\documentclass[journal = jctcce,layout=twocolumn]{achemso}

\usepackage[version=3]{mhchem} 
\usepackage[numbers,super,sort&compress]{natbib}
\usepackage[T1]{fontenc} 
\usepackage{color}
\usepackage{achemso}
\usepackage{graphicx}
\usepackage[para]{threeparttablex}
\usepackage{float}

\newcommand*{\Fig}[1]{Fig. \ref{#1}}
\newcommand*{\Tab}[1]{Tab. \ref{#1}}

\newcommand*{\CdSe}{Cd$_{33}$Se$_{33}$}

\author{O.~S.~Bokareva}
\email{olga.bokareva@uni-rostock.de}
\affiliation[Universit\"at Rostock]
{Institut f\"ur Physik, Universit\"at Rostock, Albert-Einstein-Str. 23-24, 18059 Rostock, Germany}
\author{M. F. Shibl}
\affiliation[Qatar University]
{Gas Processing Center, College of Engineering, Qatar University, P.O. Box 2713, Doha, Qatar}
\alsoaffiliation[Cairo University]
{Faculty of Science, Department of Chemistry, Cairo University, Giza, Egypt. Electronic}
\author{M. J. Al-Marri}
\affiliation[Qatar University]
{Gas Processing Center, College of Engineering, Qatar University, P.O. Box 2713, Doha, Qatar}
\author{T. Pullerits}
\affiliation[Lund University]
{Chemical Physics, Lund University, P.O. Box 124, 22100 Lund, Sweden}
\author{O.~K\"{u}hn}
\affiliation[Universit\"at Rostock]
{Institut f\"ur Physik, Universit\"at Rostock, Albert-Einstein-Str. 23-24, 18059 Rostock, Germany}

\title{Optimized long-range corrected density functionals for electronic and optical properties of bare and ligated CdSe quantum dots}
   
\begin{document}
\begin{abstract}
The reliable prediction of optical and fundamental gaps of finite size systems using density functional theory requires to account for the potential self-interaction error, which is notorious for degrading the description of charge transfer transitions. One solution is provided by parameterized long-range corrected functionals such as LC-BLYP, which can be tuned such as to describe certain properties of the particular system at hand. Here, bare and 3-mercaptoprotionic acid covered \ce{Cd33Se33} quantum dots are investigated using the optimally tuned LC-BLYP functional. The range separation parameter, which determines  the switching on of the exact exchange contribution is found to be 0.12 bohr$^{-1}$ and 0.09 bohr$^{-1}$ for the bare and covered quantum dot, respectively. It is shown that density functional optimization indeed yields optical and fundamental gaps and thus exciton binding energies, considerably different compared with standard functionals such as the popular PBE and B3LYP ones. This holds true, despite the fact that the leading transitions are localized on the quantum dot 	and do not show pronounced long-range charge transfer character.
\end{abstract}

\maketitle

\begin{figure*}[t]
	\centering
	\includegraphics[width=0.9\textwidth]{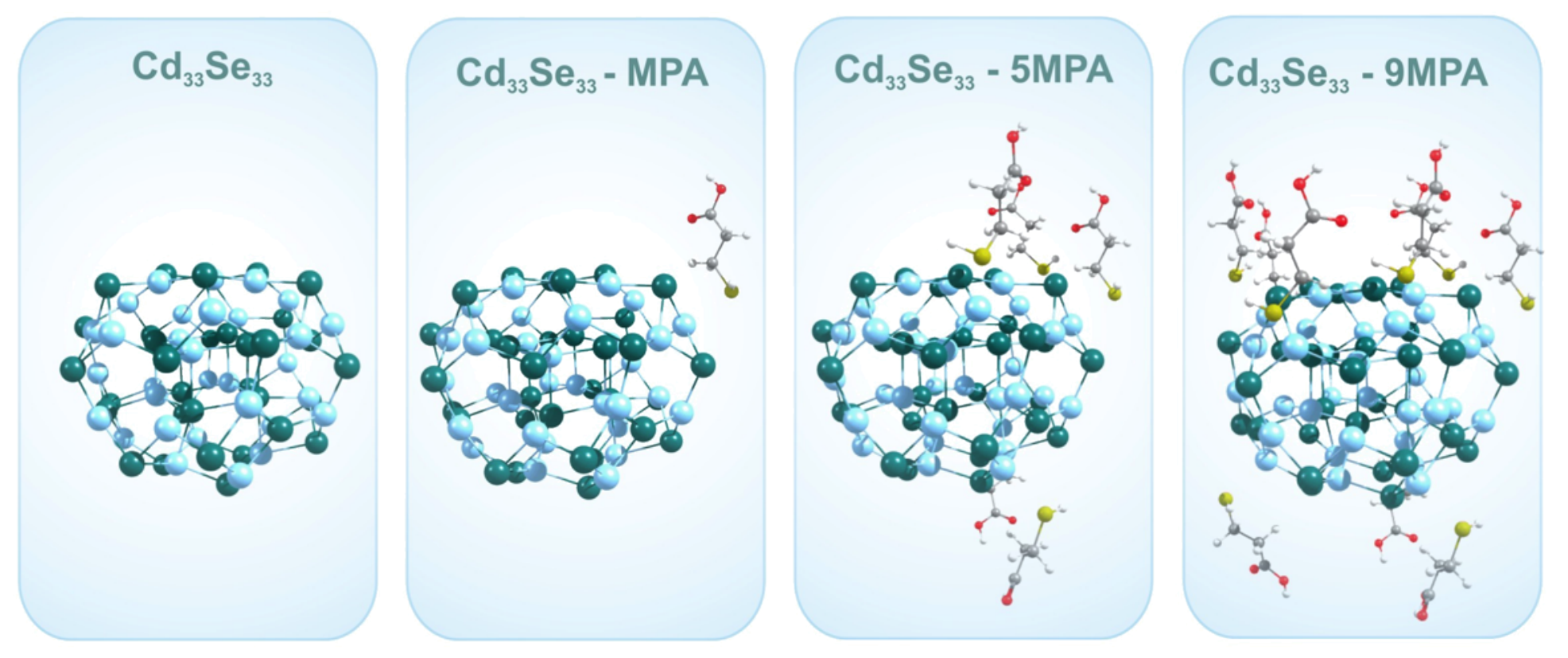}
	\caption{\label{fig:geo}
	Geometries of pure and ligated \CdSe{} QDs; the ligand is 3-mercaptoprotionic acid (MPA). The structures are taken from Ref.~\citenum{wu12}. }
\end{figure*}
\section{Introduction}
%
Semiconductor quantum dots (QDs) hold great promises for the use as tuneable light absorbers in QD solar cells (for reviews, see Refs.~\cite{scholes08_1157,kamat13_908,kovalenko15_1012}). Their flexibility derives from composition and size as well as from the control of properties by surfactants used for surface passivation during synthesis. Surface ligands are pivotal for applications such as sensing. However, in terms of electronic excitation dynamics they are discussed to play an ambivalent role. On the one hand, they suppress thermal fluctuations and thus reduce pure dephasing.\cite{liu15_9106} On the other hand, they extend the spectral density of the combined phonon and vibrational states to  significantly higher frequencies than the relatively low longitudinal optical mode frequency of the crystal. Therefore relaxation of electronic excitations due to inelastic electron-phonon scattering can be accelerated.\cite{kilina12_6515,peterson14_317,zidek14_7244}  Which effect is dominating, depends on the excitation energy with respect to the band edge. Orbitals localized on the ligands or delocalized over QD and ligands are often high in energy, although depending on the conditions and type of ligands some might be also located within the band gap.~\cite{kilina09_7717} Respective transitions are of long-range charge transfer (CT) character and, depending on the degree of ligand localization, optically weak. In this high excitation range, inelastic processes dominate, whereas for the low lying optically bright excitations elastic scattering is governing the dynamics.\cite{liu15_9106}

The nearly spherical QD, \CdSe, has gained significant attention from theory (see, e.g., Refs.~\cite{kilina09_7717,liu15_9106,kilina12_6515,abuelela12_14674,kuznetsov12_6817,trivedi15_2086}).  This cluster, which has a diameter of about 1.5~nm, is known as the ``magic'' structure, because it is the smallest one, which has a crystalline-like core. Further, it has been shown  by mass-spectroscopy to be extremely stable~\cite{kasuya04_99}.  As a model system, it possesses all features typical of QDs, most notably quantum confinement.  Its popularity among theoreticians derives from the fact that first principle calculations of static and dynamics properties can be performed with a reasonable effort.

The method of choice for simulation of QD properties is density functional theory (DFT) (for a review, see Ref.~\cite{wang15_549}). Although being a powerful method in general, the correct description of long-range CT states continues to provide a challenge for conventional as well as for hybrid functionals such as the popular B3LYP  due to the self-interaction error~\cite{dreuw03_2943,dreuw05_4009,peach08_044118}. The correct asymptotic behavior and thus a more balanced description of CT states is obtained upon introducing Hartree-Fock exchange in the long-range tail of the exchange functional. Popular variants of this idea are LC-BLYP \cite{leininger97_151,iikura01_3540}  or CAM-B3LYP \cite{yanai04_51}. In LC-BLYP, the smooth switching between short- and long-range behavior is governed by a range separation parameter. Within the so-called $\Delta$SCF method, this parameter can be obtained in a non-empirical manner, thus providing a system-specific optimization of the functional~\cite{livshits07_2932,stein09_244119,stein09_2818}. By construction, $\Delta$SCF-tuning should improve the fundamental gap and consequently all orbital-dependent properties. Successful applications have been reported for  ionization potentials (IPs), fundamental and optical gaps, CT and Rydberg excitation energies~\cite{korzdorfer14_3284,autschbach14_2592,stein10_266802,refaely-abramson13_081204}. For an application to a photocatalytic system for water splitting, see Refs.~\cite{bokareva15_1700,bokarev15_133,bokareva15_1,fischer16_404}.

Applying non-optimized CAM-B3LYP and LC-$\omega$PBE functionals to \CdSe{} and \CdSe-trimethylphosphine oxide QDs, Albert et al. concluded that both functionals severely overestimate the fundamental gap, but yield optical gaps comparable to B3LYP.~\cite{albert11_15793} The strong dependence of the optical properties on the range-separation parameter observed in Ref.~\cite{bokarev15_133} calls for reconsidering this issue in case of CdSe QDs. Therefore, in this contribution we present results for fundamental and optical gaps of the prototypical \CdSe{} bare and 3-mercaptoprotionic acid (MPA) ligated QD using a system-specific LC-BLYP functional, optimized with the $\Delta$SCF method. Compared to standard all-purpose PBE and B3LYP functionals, the optimized LC-BLYP functional yields significantly different fundamental (PBE, B3LYP) and optical (PBE) gaps. This finding, however, is in accord with experimental data.
%
\section{Computational Methods}
\label{sec:Comp}
%
As a model system we consider \CdSe{} as a bare QD and various degrees of capping with 3-mercaptoprotionic acid (MPA). It is known from   photoluminescence studies that sulfur-containing ligands tend to bind to cadmium surface atoms of QDs.~\cite{liu08_675} Full coverage with MPA is achieved by 21 ligands. In the following, at most saturation of all 9 active two-coordinated Cd atoms is considered. To investigate the effect of ligands more systematically, including also asymmetric coverages, we included the cases of one and five MPA ligands into the test set. The geometries of the considered systems are summarized in \Fig{fig:geo}. They have been taken from a study of Wu~\cite{wu12}, who performed geometry optimization using DFT with the B3LYP functional and  a LANL2DZ basis for Cd and Se as well as  a 6-31G(d) basis for the ligands. Results for a different geometry are discussed in the Supplementary Material.

In the long-range separation approach leading to the LC-BLYP functional, the Coulomb operator is split as follows:
\begin{equation}
	\frac{1}{r_{12}}=\frac{1-{\rm erf}(\omega r_{12})}{r_{12}}+ \frac{{\rm erf}(\omega r_{12})}{r_{12})}\,.
\end{equation}
Here, the parameter $\omega$ is introduced, which controls the  range of switching between the short-range (BLYP) and long-range (exact Hartree-Fock exchange) behavior by virtue of the error function. A systematic procedure for the determination of an optimal $\omega$ has been suggested in Refs.~\cite{livshits07_2932,stein09_2818,stein09_244119} and coined $\Delta$SCF method. The starting point is the observation that in exact Kohn-Sham theory the IP of the $N$-electron system equals to the negative of the energy of the HOMO. Suppose that there is some CT, the electron affinity (EA) of the latter corresponds to the IP of the $N+1$ electron system. In other words, if $E_0^{\omega}$ is the ground state energy, we have 
\begin{equation}
	{\rm IP}^{\omega}(N)= E_0^{\omega}(N-1) -E_0^{\omega}(N)\, ,
\end{equation}
which should be equal to the HOMO energy $\varepsilon^{\omega }_{\rm HOMO}(N)$, and
\begin{equation}
{\rm EA}^{\omega}(N)=	{\rm IP}^{\omega}(N+1)= E_0^{\omega}(N) -E_0^{\omega}(N+1)
\end{equation}
 which should be equal to $\varepsilon^{\omega }_{\rm HOMO}(N+1)$. If we chose the parameter $\omega$ such that the function
\begin{eqnarray} \label{Jsum} 
J(\omega ) &=& \left|\varepsilon^{\omega }_{\rm HOMO}(N)+\text{IP}^{\omega}_{}(N)\right| \nonumber\\
& + & \left|\varepsilon^{\omega }_{\rm HOMO}(N+1)+\text{IP}^{\omega}_{}(N+1)\right| 
\end{eqnarray}
is minimized, one makes a compromise for fulfilling Koopmans' theorem simultaneously for systems with $N$ (neutral complex) and  $N+1$ (anion complex)  electrons. The respective $\omega$ will be called optimal, i.e.\ $\omega_{\rm opt}$. As a note in caution, we emphasize that  since the first and second term of eq 4 refer to systems with different numbers of electrons, the minimum $J(\omega_{\rm opt})$ is not necessarily at zero, although both terms can be zero separately.  

LC-BLYP is often used with values of $\omega$, which have been obtained for certain test sets (diatomics or small molecules) and therefore are considered as standard. Typical numbers implemented in various quantum chemistry codes are $0.33 \, \text{bohr}_{}^{-1}$ and $0.47 \, \text{bohr}_{}^{-1}$~\cite{tawada04_8425,song07_154105}. There is ample evidence, however, that $\omega$-optimization for each system at hand is likely to lead to an improved performance of the LC-BLYP approach (see, e.g., Refs.~\cite{bokareva15_1700,bokarev15_133,bokareva15_1}).

Finally, we would like to point out that the thus obtained optimally-tuned functional has been critically analyzed with respect to the stability of the electronic wavefunction. This is required because of possible issues with the symmetry-breaking instabilities of ground-state solution,~\cite{cizek67_3976,bauernschmitt96_454} which may results in a divergence of TDDFT energies and an imaginary energy of the lowest triplet state ~\cite{casida00_7062}. The confirmed stability is an important prerequisite for excited states calculations reported below.

In order to characterize the QD systems we will consider the fundamental gap, $\Delta=\varepsilon^{\omega }_{\rm LUMO}-\varepsilon^{\omega }_{\rm HOMO}$, as well as the optical gap. More specifically, we will give the energies of the lowest ten transitions carrying oscillator strength. They are obtained as vertical excitation energies within the linear response formulation of time-dependent (TD-)DFT. The results will be compared with those obtained by using the standard PBE  and B3LYP functionals.~\cite{perdew96_3865,perdew97_1396} In order to characterize the QDs further and to identify CT-like transitions to the ligands, vertical excitation energies for the lowest 300 transitions have been calculated as well.

All computations have been performed using the LANL2DZ pseudopotential for Cd and Se, and the  6-31G(d) basis sets for the ligand atoms as implemented in Gaussian09~\cite{frisch09}. Pre- and post-processing of data has been done with homemade programs.

\section{Results and Discussion}
\label{sec:Res}
\subsection{Optimization of the Range-separation Parameter}

\begin{figure}[t]
	\centering
	\includegraphics[width=1.0\columnwidth]{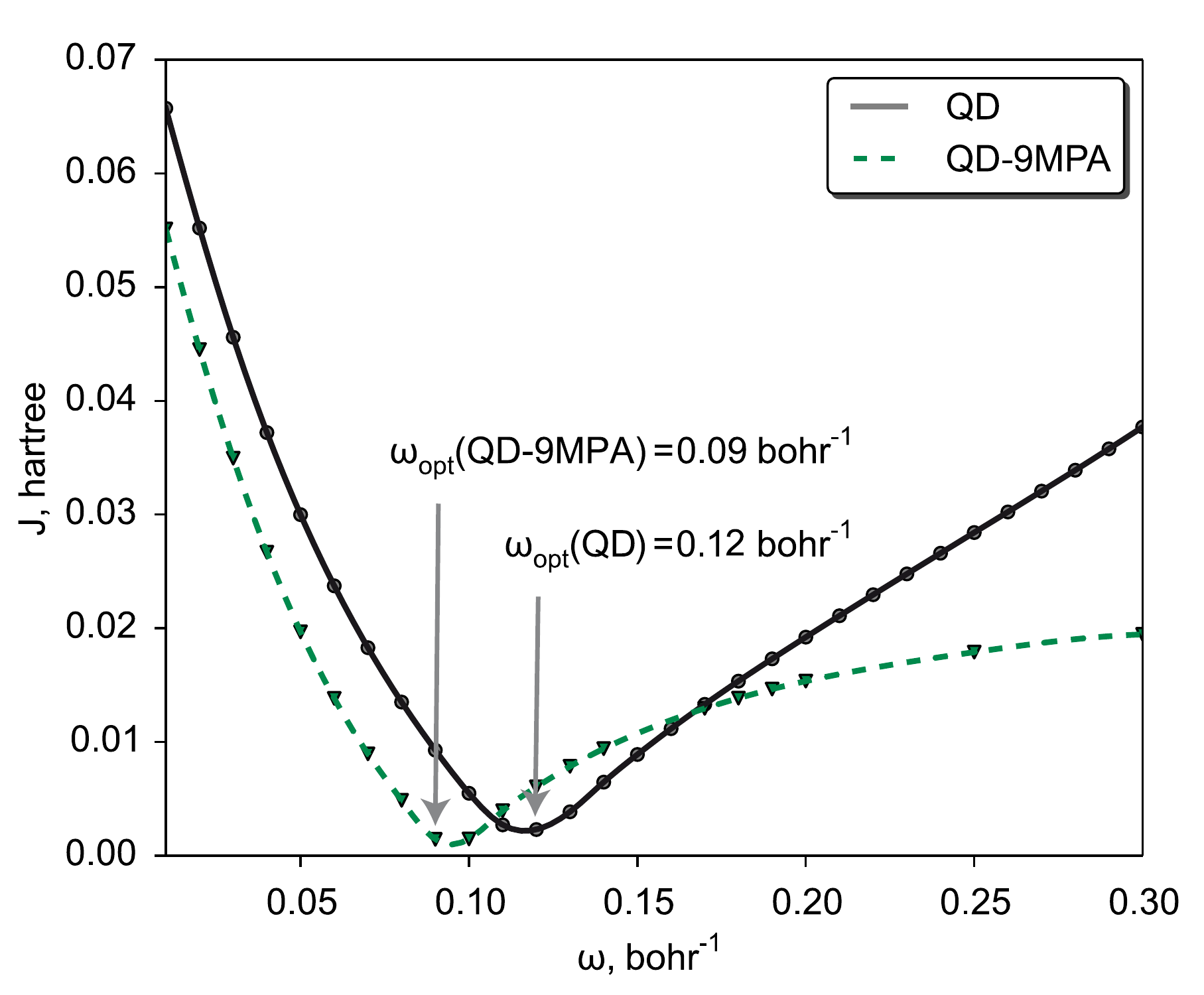}	
	\caption{\label{fig:omega}
		Dependence of the function $J$, eq 4, on the LC-BLYP range-separation parameter $\omega$ for the cases of \CdSe{} and \CdSe -9MPA.}
\end{figure}

Optimal tuning of the LC-BLYP functional using the  $\Delta$SCF method has been performed for the bare \CdSe{} and \CdSe - 9MPA, see \Fig{fig:geo}. For these cases the optimization function $J(\omega)$, eq 4,  is shown in \Fig{fig:omega}.  The closer $J(\omega)$ is to zero the better Koopmans' theorem is fulfilled. The minima at 0.12~bohr$^{-1}$ and 0.09 bohr$^{-1}$ (1~bohr=0.529~\AA) correspond to $\omega_{\rm opt}$ for \CdSe{} and \CdSe -9MPA, respectively. They are notably smaller than the standard values of 0.33 and 0.47~bohr$^{-1}$, which have been obtained by fitting to empirical data for first- to third-row atoms and to calculated atomization energies of a set of smaller molecules, respectively~\cite{tawada04_8425,song07_154105}. For both cases, one has $J(\omega_{\rm opt}) \approx 0$, i.e.\ the two conditions entering eq 4 are simultaneously fulfilled. In fact, inspecting these two conditions separately yields a rather similar $\omega$-dependence as for their sum. 

The difference between \CdSe{} and \CdSe -9MPA reflects the larger size of the latter system, which is due to the fact that $\omega^{-1}$ is a characteristic  distance for switching between short- and  long-range parts of the exchange contribution to the functional. The decrease of $\omega_{\rm opt}$  with increasing system size and  conjugation length has been found in various other applications.~\cite{stein09_244119,stein10_266802,bokareva15_1700,refaely-abramson11_075144,korzdorfer11_204107,karolewski11_151101,sears11_151103,salzner11_2568,kronik12_1515,minami13_252} Notice that this size-dependence, however, is not monotonous  and there could be a strong dependence on the electronic  structure as has been reported in Ref.~\cite{refaely-abramson11_075144}. The following results on fundamental and optical gaps have been obtained with $\omega_{\rm opt}=0.1$~bohr$^{-1}$ for all systems shown in \Fig{fig:geo}.

\subsection{Fundamental Gaps}
\label{sec:ResGr}
 
\begin{figure}[!t]
	\centering
	\includegraphics[width=1.1\columnwidth]{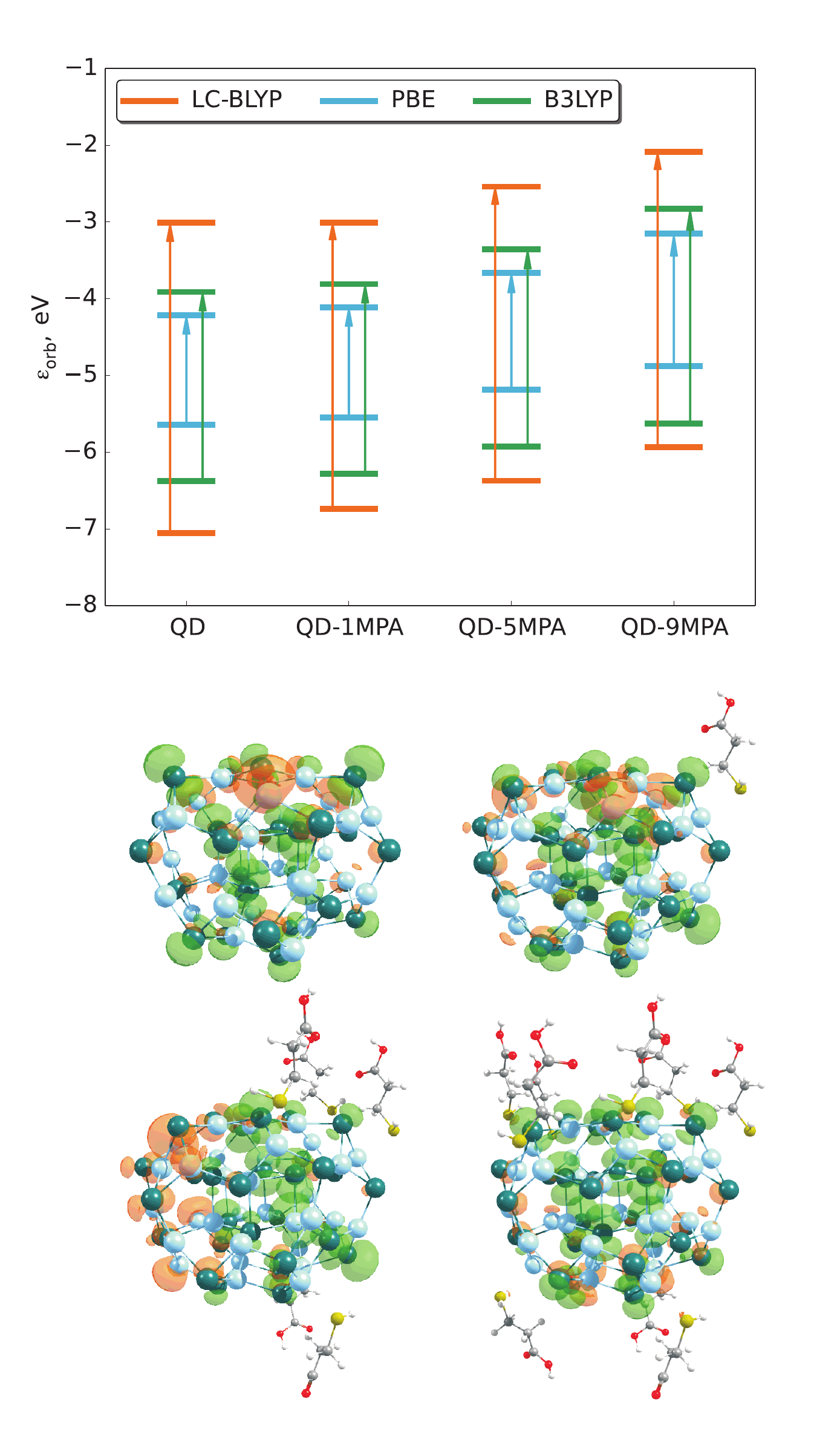}
	\caption{\label{fig:bandgap}
	Energies of HOMO and LUMO orbitals for QDs with different coverage. The results obtained with the PBE and B3LYP functionals are provided for comparison. In the lower panel electron density differences (contour value 0.0003, orange - particle, green - hole) are shown, corresponding to the HOMO-LUMO excitation within LC-BLYP. }
\end{figure}

The optimally-tuned LC-BLYP functional has been used to obtain ground state energies and molecular orbitals for the systems shown in \Fig{fig:geo}. In the following,	 fundamental gap energies, $\Delta$, are compared with those calculated by using the PBE and B3LYP functionals. The results are summarized in \Fig{fig:bandgap} and \Tab{tab:bandgap}. Comparing the HOMO and LUMO energies obtained with the optimized LC-BLYP functional for different QD coverages, the following trends can be deduced. First, upon increasing the number of ligands on the surface, the HOMO and LUMO energies increase. The fundamental gap is $\Delta=4.04$~eV for the bare QD and $\approx 3.8$~eV in the presence of the ligands. The number of ligands has only little influence on $\Delta$.

In the bottom part of \Fig{fig:bandgap}, the LC-BLYP electron density differences for the HOMO-LUMO transitions  for all studied complexes are presented. All frontier orbitals in the systems are localized purely on the QD, the percentage of electron density on the QD is always more than 95\%. Although the absolute orbital energies differ notably for the applied functionals, the density differences for the HOMO-LUMO transition  are very similar for  PBE and LC-BLYP. Ligation with a single MPA almost does not change the electron density difference upon the HOMO-LUMO transition of the bare QD. The symmetrical coverage with 9 MPAs leads to minor changes in the  hole (green color) density but the particle (orange) density is redistributed. In all three above-mentioned cases, the transitions show only slight polarization, i.e.\ CT-like character. The only case with notable polarization appears for the QD, partially covered with 5 MPAs.

In passing we note that the effect of hydrocarbon chain length elongation on the HOMO-LUMO properties has been studied for the example of QD-1MPA. Here,  up to  ten additional CH$_2$-groups have been added. The shapes of HOMO and LUMO are essentially not affected and the changes in the orbitals energies and $\Delta$ are rather small (about  -0.01~eV for most cases).

\begin{table*}[t]
\begin{center}
\begin{threeparttable}
\caption{\label{tab:bandgap}
	Values of the fundamental gap,  $\Delta$, (in eV) of QDs with different coverage obtained by using different DFT functionals (left three columns with B3LYP geometry of Ref.~\cite{wu12}).}
\begin{tabular*}{0.9\textwidth}{@{\extracolsep{\fill}} l | c c c c c}
\hline
				& LC-BLYP &  PBE & B3LYP & B3LYP\cite{kilina09_7717, kuznetsov14_7094, wu12} & various~\cite{kuznetsov14_7094} \\
				\hline
				\CdSe & 4.04 & 1.43 & 2.46 & 2.42-2.80$^a$& 1.39-2.89$^b$\\
				\CdSe -MPA & 3.73 & 1.44 & 2.47 &&\\
				\CdSe -5MPA & 3.83 & 1.53 & 2.57 &&\\
				\CdSe -9MPA & 3.85 & 1.73 & 2.80 & 2.70-2.99$^c$&\\
				\hline
\end{tabular*}
\begin{tablenotes}
\footnotesize
\item[a] Value depends on the geometry of the QD;
\item[b] $\Delta$ increases according to LSDA, PBE, PW91, B3LYP, PBE0;
\item[c] for QD ligated with 9 methylamine or trimethylphosphine oxide for different geometries.
\end{tablenotes}
\end{threeparttable}
\end{center}
\end{table*}
Comparing  optimized LC-BLYP and PBE results one notices that in the latter case $\Delta$ is predicted to be substantially smaller (on average by a factor of 2.5). This underlines the importance of application of long-range separated functionals even for cases where the relevant orbitals are localized solely on the QD. As can be seen from \Fig{fig:bandgap} and \Tab{tab:bandgap}, PBE reproduces the general dependence of HOMO and LUMO energies on the QD coverage. However, it gives a too small value of $\Delta$ because of an overestimation of $\varepsilon_{\rm HOMO}$ and underestimation of $\varepsilon_{\rm LUMO}$. Moreover, for PBE the trend concerning the dependence of $\Delta$ on the coverage is opposite to that obtained for LC-BLYP.

It is well-established that fundamental gaps of QDs are sensitive to the amount of exact exchange in the functional.  In Ref.~\cite{kuznetsov14_7094} a comparative analysis of  $\Delta$  using different functionals (B3LYP, LSDA, PBE, PW91, PBE0) has been performed for \CdSe{} and  Cd$_{33}$Te$_{33}$ QDs (cf. Suppl. Mat.). As a conclusion, the B3LYP functional, which has a constant portion of exact Hartree-Fock exchange (20\%), has been recommended. Values for $\Delta$ for the present case and those available from literature are summarized in \Tab{tab:bandgap}. Note that there is a certain dependence on the geometry of the QD. In Ref.~\cite{kilina09_7717} slight variations of geometry have been shown to lead to changes in $\Delta$ of about 15\%.  Clearly, although B3LYP gives an improvement as compared to PBE, the fundamental gap is still considerably underestimated with respect to LC-BLYP (by an average factor of 1.5 if compared to a value of 2.5 for PBE). 

\subsection{Optical Gaps}

\begin{figure*}[!t]
	\includegraphics[width=1.0\textwidth]{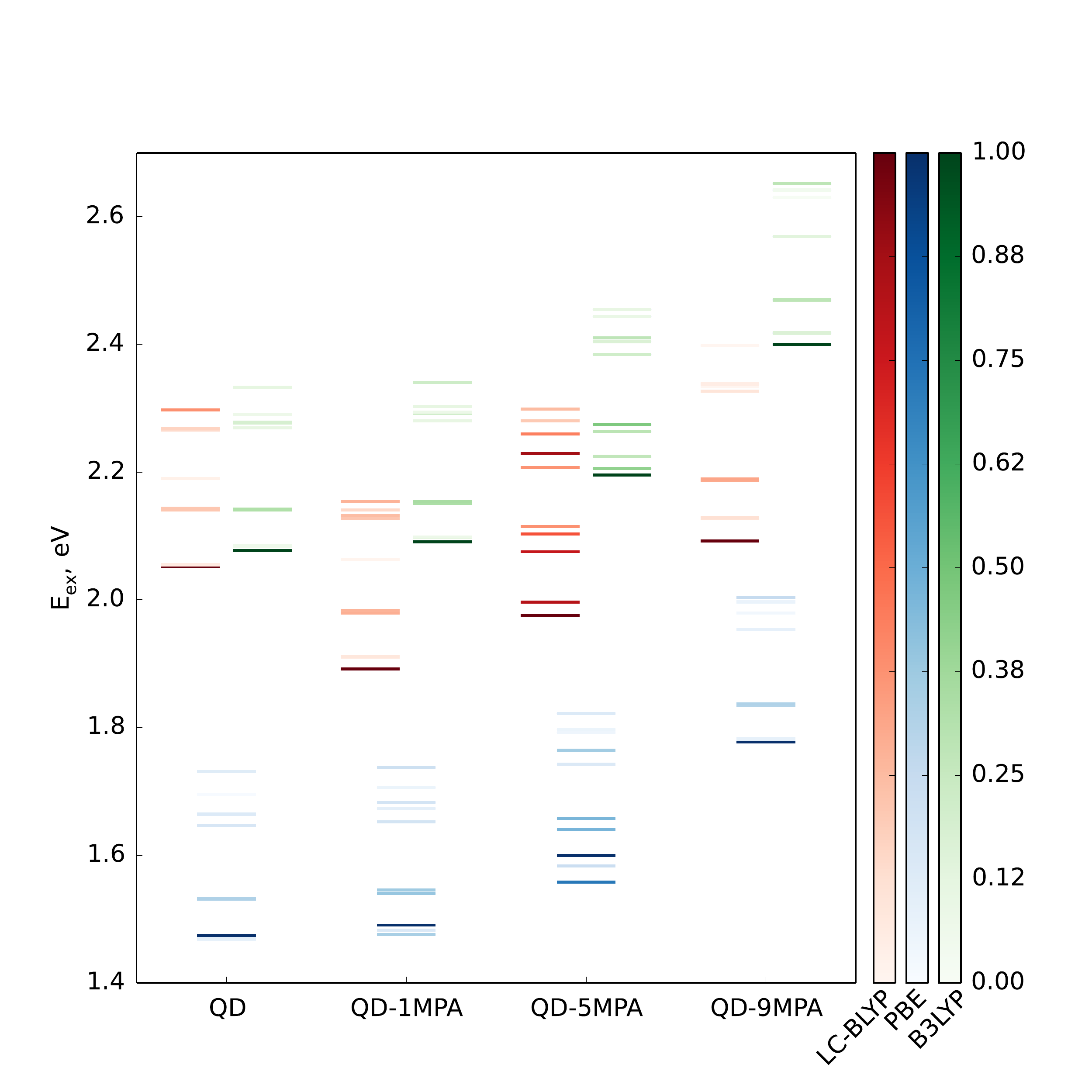}
	\caption{\label{fig:opticalgap}
	Optical gaps for the systems shown in \Fig{fig:geo} as obtained by using the optimized LC-BLYP, PBE, and B3LYP functionals. Shown are the lowest ten transitions with the color code corresponding to the oscillator strength (normalized separately for each case). }
\end{figure*}

\Fig{fig:opticalgap} summarizes the LC-BLYP results obtained for the optical gaps (actually, the ten lowest transitions, a wider range of the spectrum is discussed below in \Fig{fig:spectra}) for the different systems. For the bare QD the lowest transition around 2.05~eV. The effect of MPA depends on the coordination of the Cd surface atoms where the ligands are attached to. Two- and three-coordinated Cd sites tend to yield a blue and red shift, respectively, of the optical gap (see also Ref.~\cite{kilina09_7717}, where the spectra of QD covered with different numbers of amine and phosphine oxide ligands have been compared). In the present case, binding to two-coordinated site is considered only. The addition of the first MPA-ligand shifts the optical gap to the red. The addition of further ligands  yields a blue shift.  Analyzing the transitions for the present system, we note that in all cases the leading contributions in this energy range involve orbitals, which are localized on the QD. 

\Fig{fig:opticalgap} also contains the comparison with PBE and B3LYP results. Not surprisingly for PBE the optical gaps are substantially lower than for LC-BLYP.  The strongest transition is predicted to be at 1.47~eV, i.e.\ well below the experimental value. Further, the red shift of the optical gap for the \CdSe -1MPA is not observed, i.e.\ there is almost no effect due to the addition of a single ligand. The blue shift with increasing coverage is reproduced. B3LYP shows the same trend as PBE as far as the dependence on coverage is concerned. However, the absolute values are in general above the LC-BLYP ones. Only in case of the bare QD LC-BLYP and B3LYP yield almost identical optical gaps. Overall, the B3LYP results are closer to the LC-BLYP ones  (average deviation 10\%)  as compared to the PBE ones (average deviation 12\%). 

\begin{table*}[t]
\begin{center}
\begin{threeparttable}
\caption{\label{tab:mixing}
	The optical gap energies and the oscillator strengths (in parentheses)  for the systems shown in \Fig{fig:geo}.  Also given are the  characters of the transitions  in terms of the leading orbitals (with respect to the HOMO (H) and LUMO (L)).}
\begin{tabular*}{1.0\textwidth} {l | c c| c  c| c  c}
\hline
&  LC-BLYP &  & PBE &  & B3LYP &\\
\hline
 & energy (eV) & character & energy (eV) & character & energy (eV) & character \\
 \hline
\CdSe  &  2.05 (0.092) & 38\% H-L & 1.47 (0.003) & 33\% (H-1)-L & 2.08 (0.067) & 25\% (H-1)-L \\
&&&& 16\% H-L && 18\% H-L \\
-MPA &   1.89 (0.075) &35\% H-L &   1.48 (0.011) & 47\% H-L & 2.09 (0.063) & 25\% H-L \\
&&&& &&13\% (H-1)-L \\
-5MPA &  1.98 (0.040) &15\% (H-4)-L &   1.56 (0.019) & 49\% H-L & 2.20 (0.052) &27\% H-L  \\ 
&& 13\% (H-1)-L   &&&&19\% (H-2)-L\\
-9MPA &  2.09 (0.110) &46\% H-L &   1.77 (0.069) & 49\% H-L & 2.40 (0.107) & 48\% H-L \\ 
 \hline
\end{tabular*}
\end{threeparttable}
\end{center}
\end{table*}

Previously, it has been pointed out that the excitonic character of the lowest transitions manifests itself in their strongly mixed character in terms of single particle molecular orbitals~\cite{delpuerto06_096401}. In Ref.~\cite{delpuerto06_096401} this has been demonstrated most clearly by comparing results of DFT calculations in the local density approximation with  solutions of the Bethe-Salpeter equation in GW-approximation for nonstoichiometric CdSe clusters. \Tab{tab:mixing} summarizes the character of the optical gap transition for the optimized LC-BLYP, PBE, and B3LYP functionals. First, we notice that the character of the transitions is mixed in all cases, i.e.\ the leading contributions in terms of the single particle orbitals are below 50~\%. Second, LC-BLYP predicts a stronger mixing if compared to PBE and B3LYP. Finally, the degree of mixing is notably influenced by the coverage with the MPA ligand, although there is no obvious trend.

\subsection{Electronic Excitation Spectra} 

\begin{figure}[t!]
	\centering
	\includegraphics[width=1.0\columnwidth]{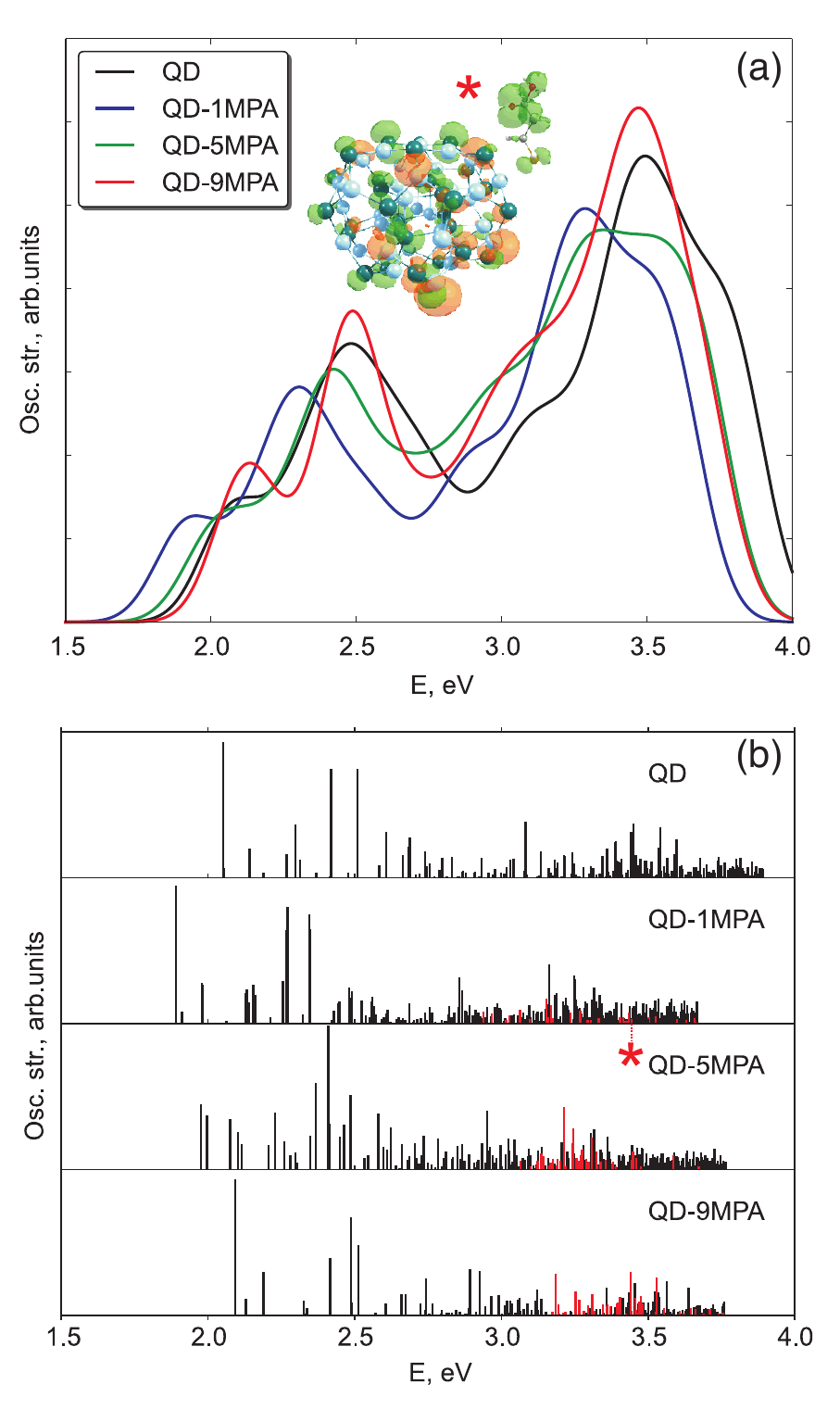}
	\caption{\label{fig:spectra}
		(a) Broadened absorption spectra for the systems shown in \Fig{fig:geo} as calculated with the tuned LC-BLYP functional. (b) Absorption stick spectra (normalized separately). Red sticks denote transitions with admixture of CT character. An exemplary density difference of a  CT-like transition of QD-1MPA (marked by the  asterisk) is shown in panel (a).}
\end{figure}

The experimental absorption spectra of CdSe QDs are not very structured and the most notable feature is a peak at the  onset of the spectra, which shifts to lower energies with increasing QD size~\cite{scholes08_1157}.  For the particular case of \CdSe{} with decylamine surfactants, this peak has been reported at about 3~eV (415~nm) in toluene solution~\cite{kasuya04_99}. \Fig{fig:spectra}a compares the broadened absorption spectra of the different system up to about 1.5~eV above the optical gap. In principle, the peak at the onset of the spectrum is clearly discernible. Its position depends strongly on the coverage, what hampers comparison with available experimental data.

Ligands such as MPA may in principle introduce trap states at the surface of the QD, i.e.\ states localized at the ligands. Transitions to such trap states will be of CT type, i.e.\ their proper description could be very sensitive to the functional and using a system-specific long-range corrected functional will be beneficial. For the  case of \CdSe{} covered with various numbers of OPMe$_3$ or NH$_2$Me-9MPA, B3LYP predicts CT transitions about 2~eV above the optical band gap~\cite{kilina09_7717}. In \Fig{fig:spectra}b stick spectra of the different systems are shown and  transition being of notable CT-character are highlighted. The degree of CT can be judged from the density difference plot given for a particularly strong CT-like transition.  For the considered systems the onset of absorption is slightly dependent on the coverage. However, this is an electronic effect  of the ligands, which doesn't seem to be related to the appearance of low-lying states with pronounced CT character. In fact partly CT- like transitions are located more than 1~eV above the onset of the spectrum.

%

\section{Discussion and Conclusions}
\label{sec:Conc}
%

Optimization of the long-range corrected LC-BLYP functional has been applied to study the fundamental and optical gaps of bare and covered \CdSe{} QDs. Although for the considered systems the frontier orbitals are mostly localized on the QD, the effects of intra-QD charge rearrangement when going from the HOMO to the LUMO are substantial. In particular fundamental gap energies  for PBE and B3LYP are severely underestimated if compared to the LC-BLYP optimized functional.  For the optical gap, results for  the PBE and the LC-BLYP optimized functional also differ although not that markedly. In general, the B3LYP results are closer to the LC-BLYP ones  as compared to the PBE ones. For the bare QD LC-BLYP and B3LYP give almost identical optical gaps. This good performance of B3LYP  is in accord with previous investigations.~\cite{kilina09_7717} 

Comparing fundamental and optical gaps one obtains the exciton binding energy. For the bare QD this yields 1.99 eV (LC-BLYP), -0.04 eV (PBE), and 0.44 eV (B3LYP) and for \CdSe -9MPA 1.67 eV (LC-BLYP), -0.04 eV (PBE), and 0.39 eV (B3LYP). Leaving aside the apparently erroneous PBE values, LC-BLYP exciton binding energies are considerably larger than B3LYP ones. Unfortunately, there are no experimental data for \CdSe. However, combining X-ray absorption and photoelectron spectroscopy for the determination of the fundamental gap with optical absorption spectroscopy, the exciton binding energy for CdSe QDs in hexane of about the same size has been determined to be about 1~eV.~\cite{meulenberg09_325}  Unfortunately, this neither agrees with the B3LYP not with LC-BLYP result. According to Ref.~\cite{meulenberg09_325} the fundamental gap should be around 4.3~eV, i.e. confirming the present LC-BLYP value. The strong absorption peak at the onset of the spectrum is found at about 3.3~eV in Ref.~\cite{meulenberg09_325} and 3~eV in Ref.~\cite{kasuya04_99}. Hence we conclude that both, LC-BLYP (2.05~eV) and B3LYP (2.08~eV) underestimate this transition and thus give different values for the exciton binding energy. There could be at least two reasons for this discrepancy in case of LC-BLYP. First, the optimization of the range-separation parameter sets the focus on the HOMO-LUMO gap. Thus the functional is not optimal for the present optical transitions, which are of rather mixed character. Second, there appears to be a rather strong dependence on geometry (cf. Suppl. Mat.). The present B3LYP optimized geometry of the bare QD might differ from the LC-BLYP one, but also from the actual experimental geometries under solution phase conditions.

To conclude, for a balanced treatment of fundamental and optical gaps, exact Hartree-Fock exchange should be included via optimized range-separated functionals such as LC-BLYP. This issue will be even more pressing once orbitals localized on the ligands need to be considered. Given the poor performance of non-optimized range-separated functionals in Ref.~\cite{albert11_15793}, system-specific functional optimization appears to be mandatory.  For the present case of bare and covered \CdSe{} QDs, an LC-BLYP range separation parameter of 0.1~bohr$^{-1}$ is recommended.
\begin{acknowledgement}
This work was made possible by NPRP grant \#NPRP 7-227-1-034 from the Qatar National Research Fund (a member of Qatar Foundation). The statements made herein are solely the responsibility of the authors.
\end{acknowledgement}

\begin{suppinfo}
	Comparison of fundamental and optical gaps for different geometries \CdSe. 
\end{suppinfo}

\providecommand{\latin}[1]{#1}
\providecommand*\mcitethebibliography{\thebibliography}
\csname @ifundefined\endcsname{endmcitethebibliography}
  {\let\endmcitethebibliography\endthebibliography}{}

\end{document}